# Magneto-ionic Control of Interfacial Magnetism

Uwe Bauer[1], Lide Yao[2], Satoru Emori[1], Harry L. Tuller[1], Sebastiaan van Dijken[2] and Geoffrey S. D. Beach[1*]

[1]Department of Materials Science and Engineering, Massachusetts Institute of Technology, Cambridge, Massachusetts 02139, USA

[2]Department of Applied Physics, Aalto University, P.O. Box 15100, FI-00076 Aalto, Finland

\* Author to whom correspondence should be addressed. Electronic mail: gbeach@mit.edu, Phone: +1 617 258-0804

In metal/oxide heterostructures, rich chemical,[1,2] electronic,[3-5] magnetic[6-9] and mechanical[10,11] properties can emerge from interfacial chemistry and structure. The possibility to dynamically control interface characteristics with an electric field paves the way towards voltage control of these properties in solid-state devices. Here we show that electrical switching of the interfacial oxidation state allows for voltage control of magnetic properties to an extent never before achieved through conventional magnetoelectric coupling mechanisms. We directly observe, for the first time, *in situ* voltage driven $O^{2-}$ migration in a Co/metal-oxide bilayer, which we use to toggle the interfacial magnetic anisotropy energy by >0.6 erg/cm$^2$. We exploit the thermally-activated nature of ion migration to dramatically increase the switching efficiency and to demonstrate reversible patterning of magnetic properties through local activation of ionic migration. These results suggest a path towards voltage-programmable materials based on solid-state switching of interface oxygen chemistry.

The physical and chemical properties of nanoscale materials derive largely from structure and composition at interfaces. The possibility to electrically modify these interfacial characteristics would provide a powerful means to control material properties. Of particular recent scientific and technological interest are metal/metal-oxide bilayers,[1-15] in which properties as varied as catalytic activity,[1,2] charge and spin transport,[3-6] ionic exchange,[14,15] mechanical behavior,[10,11] thermal conductivity[12,13] and magnetism[6-9] all depend sensitively on oxygen stoichiometry and defect structure at the metal/metal-oxide interface. Ionic transport in metal-oxides can be driven by an electric field, and $O^{2-}$ migration is already exploited as a mechanism for resistive switching in anionic metal/oxide/metal memristors[3,4]. However, the broader application of voltage-driven oxygen transport to control interfacial phenomena in metal/metal-oxide structures is only beginning to be explored.

For example, ferromagnetic metal/metal-oxide bilayers such as Co/AlO$_x$ and CoFe/MgO exhibit strong perpendicular magnetic anisotropy (PMA) derived from interfacial Co-O hybridization.[8,9] The ability to toggle interfacial PMA with a gate voltage would dramatically reduce switching energies in spintronic devices, and could enable new device architectures exploiting local gating of magnetic properties.[16-26] Most work on voltage control of magnetic anisotropy in metal/metal-oxide bilayers has focused on charge accumulation or band shifting in the metal layer.[16-19] However, experimental reports of irreversibility, and anisotropy changes much larger than theoretically predicted,[20,21] suggest ionic effects may be important and in some cases even dominant.[22-26] Nonetheless, electric field driven oxygen migration in metal/metal-oxide bilayers is difficult to observe directly, and the impact on magnetic properties has not yet been established.

Here we report direct *in-situ* observation of voltage-driven $O^{2-}$ migration in a metal/metal-oxide bilayer, and show that solid-state electro-chemical switching of the interfacial oxidation state can be used to completely remove and restore PMA in a thin Co layer. Using cross-sectional transmission electron microscopy (TEM) and high-resolution electron energy loss spectroscopy (EELS), we track *in-situ* voltage-driven migration of the oxidation front in a Co/GdO$_x$ bilayer. By varying temperature and interface structure, we relate motion of the oxidation front to voltage-induced anisotropy changes, and show that increasing the temperature by just ~100°C above ambient reduces the timescale of these effects by ~4 orders of magnitude. We toggle interfacial PMA by an unprecedented >0.6 erg/cm$^2$ at just 4V, yielding a magnetoelectric efficiency >5000 fJ/Vm, establishing magneto-ionic coupling as a powerful mechanism for voltage control of magnetism. Finally, we introduce a general method to reversibly imprint material properties through local activation of ionic migration, which we use to locally pattern magnetic anisotropy and create domain wall conduits in continuous magnetic films.

Experiments focus on Ta(4 nm)/Pt(3 nm)/Co(0.9 nm)/GdO$_x$(3 nm) films sputter-deposited on thermally-oxidized Si (see Methods). The films exhibit strong PMA with an in-plane saturation field $H_k$~10 kOe (see Fig. S3). Square 200μm x 200μm GdO$_x$(30 nm)/Ta/Au gate electrodes were patterned on top of the film for voltage application, with the bottom metal stack used as counter electrode. Fig. 1a shows a high-resolution cross-sectional TEM micrograph of the layer structure, with the thin Co layer embedded between polycrystalline Pt and GdO$_x$ layers.

Spatially resolved EELS experiments were carried out in scanning TEM (STEM) mode on the same cross section to measure the chemical profile and to detect changes induced by a gate voltage $V_g$. Voltage was applied by contacting the top electrode *in situ* with a Pt/Ir tip as shown in Fig. 1b (see SI for details). Fig. 1c shows representative STEM-EELS spectra with O K-edges and Co white lines from a location in the center of the Co layer. The evolution of the O K-edge and Co $L_3$-edge count rates along a line profile perpendicular to the layers is shown in Fig. 1d. The Co layer is clearly distinguished, with no O detected within the Co layer and a sharp rise in the O signal at the Co/GdO$_x$ interface. Figs. 1e,f show STEM-EELS scans after applying a negative bias of -3V, and -5V, respectively, each for ~100s, with a polarity that drives $O^{2-}$ towards the Co layer. The data show a clear progression of the oxygen front into the Co layer, with a significant O signal detected near the center of the Co layer after -3V, and at the bottom Co/Pt interface after -5V. The appearance of an O K-edge signal at the center of the Co layer after the application of -3V can also be inferred from the spectra in Fig. 1c.

We examined the impact of voltage-induced $O^{2-}$ migration on magnetic properties using a scanning magneto-optical Kerr effect (MOKE) polarimeter with a ~3 µm laser spot to locally probe hysteresis characteristics. Fig. 2a maps the coercivity $H_c$ in the vicinity of a gate electrode, shown schematically in Fig. 2b. Prior to measurement, a domain nucleation site was created nearby using a mechanical microprobe[22,23,27] so that $H_c$ represents the domain wall (DW) propagation field,[22,23,27] which is highly sensitive to the magnetic anisotropy energy (MAE) landscape.

In the virgin state (Fig. 2a), $H_c$ is uniform across the measured area, reflecting a DW propagation field of ~200 Oe due to fine-scale disorder. After applying $V_g$ = -4V for 240s and then setting $V_g$=0V, $H_c$ exhibits an abrupt step at the electrode edge (Fig. 2c) and increases to ~340 Oe beneath the electrode. Similar behavior has been observed previously[23] and attributed to $O^{2-}$ migration near the electrode perimeter, where ionic transport is typically most efficient[28]. This would locally reduce the MAE energy by over-oxidizing the Co interface[8,9], creating potential wells at the electrode edge (Fig. 2d) that trap propagating DWs, increasing $H_c$. We verified that the DW propagation field at the interior of the electrode remained unchanged after $V_g$ application by mechanically creating a nucleation site inside the electrode. The observed magnetic behavior is thus consistent with the schematic MAE landscape in Fig. 2d.

The lack of irreversible MAE changes at the electrode interior suggests the timescale for bulk $O^{2-}$ diffusion is much longer than at the electrode perimeter, where the open oxide edge (Fig. 2b) provides a high-diffusivity path[28]. The high ionic mobility observed in Fig. 1 is likewise probably aided by the high surface to volume ratio of the polished TEM specimen, since the activation energy for surface diffusion is typically lower than for bulk[28]. Voltage-induced $O^{2-}$ migration, however, need not be limited to the oxide edge. Due to the thermally-activated nature of ion migration, voltage application at elevated temperature should result in exponentially higher $O^{2-}$ drift velocities,[28] and activation of bulk $O^{2-}$ migration on an observable timescale.

Figure 2e shows a $H_c$ map after applying $V_g = -4V$ for 155s at $T=100^oC$, and then cooling the sample back to room temperature at $V_g=0V$. In this case, $H_c$ outside the electrode is unchanged, showing that the elevated temperature alone does not irreversibly change the magnetic properties (see Fig. S4). However, beneath the electrode $H_c$ drops to ~50 Oe, indicating that DWs nucleate and propagate there at a much lower field, and are impeded by an anisotropy step at the electrode edge. This implies a significant MAE reduction across the electrode area, shown schematically in Fig. 2f, consistent with overoxidation of the Co interface [8,9].

As seen in Figs. 2g-k, the voltage-induced MAE change at $T=100^oC$ is progressive. Here, we sequentially applied $V_g = -4V$ for a fixed dwell time at $T=100^oC$, and then cooled the sample to room temperature with $V_g=0V$ to measure a hysteresis loop beneath the electrode. We observe an initial increase in $H_c$ due to DW trap formation at the electrode perimeter, which occurs within 1s of $V_g$ application in contrast to the several minutes required at room temperature (Fig. 2c). With increasing voltage dwell time $H_c$ then drops dramatically, indicating a rapid reduction of PMA across the electrode. The saturation MOKE signal also declines, by nearly a factor of 2 after ~150s (Fig. 2i), suggesting increasing Co oxidation. After several minutes (Fig. 2j), PMA is lost entirely. Remarkably, PMA can be completely restored by reversing the bias polarity, as seen in Fig. 2k after applying $V_g$=+4V for 270s at 100°C.

To correlate magnetic anisotropy with the location of the oxidation front, we used a Gd spacer to control the distance between Co and the Gd/GdO$_x$ interface. Here, GdO$_x$ was grown as a continuous 30nm-thick layer before depositing Ta/Au electrodes to prevent DW trap formation at the edges. Figs. 3a-d show the nominal sample structure and corresponding hysteresis loops for four samples with increasing Gd spacer thickness $d$. With a thin Gd spacer, PMA is diminished as evidenced by a significant drop in the remanent magnetization ratio $M_r/M_s$ (Fig. 3b), but as $d$ is increased further, PMA again increases (Figs. 3c-d). This non-monotonic anisotropy transition, different from the behavior reported for Pt/Co/Al/AlO$_x$,[8,9] was also observed for other metal-oxides such as Pt/Co/Zr/ZrO$_x$. Although its origin is currently unclear, the non-monotonic behavior is fortuitous because it allows us to determine sensitively the position and direction of motion of the oxidation front.

Figs. 3e-h show that $V_g$ applied to the sample in Fig. 3c completely reproduces the non-monotonic anisotropy transition exhibited by the as-deposited samples (Fig. 3a-d). With $V_g$>0, PMA gradually increases (Fig. 3h) whereas with $V_g$<0, PMA first decreases (Fig. 3f) then slowly increases with increasing dwell time (Fig. 3e) (See also Fig. S5). The effect of positive (negative) $V_g$ is thus equivalent to increasing (decreasing) the spacing between Co and the oxidation front.

Finally, we show in Fig. 3f that increasing $T$ and $V_g$, to 120°C and 12 V, respectively, decreases the timescale for anisotropy switching to <10ms, compared to ~100s at 100°C and 6V, and in contrast to the inaccessibly-long timescales required at room temperature. The results in Figs. 2 and 3 demonstrate that these dramatic, nonvolatile MAE changes occur through a thermally-activated process, consistent with voltage-induced $O^{2-}$ migration as observed in Fig. 1. The magnitude of the effect, which yields in Figs. 2g-k a change in interfacial PMA >0.6 erg/cm$^2$, or ~5000 fJ/Vm, corresponds to the largest magneto-electric coupling efficiency yet reported.

Based on these findings we demonstrate local MAE patterning using the MOKE laser spot to locally heat the sample and activate oxygen migration under $V_g$. Using the devices in Fig. 2b, we first apply $V_g$<0 at room temperature to create a potential well at the electrode perimeter, evidenced by a jump in $H_c$ beneath the electrode (Fig. 4a). This isolates the electrode area from DW motion in the adjacent Co film. With $V_g$ applied, we then increase the MOKE laser power $P$ from 1mW to 10mW, corresponding to a local temperature rise of ~20°C (see Methods). This causes $H_c$ and $M_r/M_s$ to drop dramatically underneath the spot (Fig. 4a), and this change is retained when $V_g$ is removed and $P$ decreased to 1mW. Neither $V_g$ nor high $P$ alone is sufficient to cause irreversible changes, but when both are sufficiently high (see Fig. S8), laser-induced heating activates voltage-driven $O^{2-}$ migration, facilitating local MAE imprinting.

Due to the local anisotropy reduction, the illuminated spot acts as a domain nucleation site, as seen in the time-resolved scanning MOKE images in Figs. 4b-d (see Methods). To demonstrate the reversibility of this MAE imprinting, we first created a laser-induced DW nucleation site in one corner of an electrode and then placed the laser spot in the diagonally opposite corner (Fig. 4e) with $P$ = 10 mW. At $V_g$=0 the higher laser power alone has no effect on $H_c$, which is determined by the field necessary to propagate a DW from the far corner. But with $V_g$=-3 V, $H_c$ and $M_r/M_s$ drop dramatically, indicating that instead of propagating across the electrode, DWs nucleate directly underneath the laser spot due to the local PMA reduction. Positive bias restores $M_r/M_s$ and $H_c$ to their initial values and $H_c$ can be toggled repeatedly in this manner as $V_g$ is cycled between +3V and -3V ( Fig. 4f).

Finally, we imprint more complex anisotropy patterns that allow for spatial control of magnetization dynamics. In Fig. 5, we define a conduit in which DWs are injected from a laser-written nucleation site and propagate along a pre-defined path. At $V_g$ = -3 V, a point inside the electrode area was illuminated at $P$=10mW for 100 s to produce a DW nucleation site. The laser was then scanned along an L-shaped line in 1.25 µm steps with a variable dwell time to produce a conduit of reduced MAE (Fig. 5a). Figs. 5a–f show time-resolved MOKE images of field-driven domain expansion in the patterned region (see Methods). Here, a reverse domain nucleates at the laser-defined nucleation site (Fig. 5b) and expands preferentially along the laser-written conduit (Figs. 5c-f). The degree of confinement depends on the difference in MAE in the film and in the patterned region, which determines the difference in creep velocity along and orthogonal to the conduit. The DW velocity follows $v \propto \exp(-E_a(H)/k_BT)$, where the activation energy $E_a(H) \propto H^{-1/4}$ depends on the MAE.[27] Fig. 5g shows that the slope of $\ln(v)$ versus $H^{-1/4}$, and hence the activation energy that determines the DW velocity, can be precisely tuned to control the DW dynamics. By reducing the anisotropy in the conduit we enhance the velocity by up to a factor of ~160. The MAE can be further reduced, but in this case nucleation along the conduit is observed.

We note that local MAE patterning has previously been demonstrated using local ion beam irradiation,[29,30] but has never been realized in a nondestructive and completely reversible way.

Here, the spatial resolution is limited by the laser spot size to a few μm. However, this resolution limit could easily be overcome by instead heating the sample globally and writing the anisotropy pattern locally using, e.g. a conductive AFM tip to apply $V_g$ with high spatial resolution.

Our work shows interfacial chemistry in metal/metal-oxide bilayers can be electrically-gated using an all-solid-state device, operating at low voltage and within the typical operating temperature range of common semiconductor electronics. Specifically for Co/metal-oxide bilayers, where interfacial MAE is sensitive to interface oxygen coordination, we use voltage control of oxygen stoichiometry to achieve unprecedented control over magnetic anisotropy. Moreover, we show that relatively small changes in temperature and gate voltage can improve device response times by orders of magnitude, so that considerable further improvement can reasonably be anticipated using materials with even a modestly lower energy barrier for ionic migration. Although this work focused on magnetic properties, reversible voltage-gated control of oxygen stoichiometry in metal/metal-oxide bilayers makes a wide range of materials properties and effects amenable to solid-state electrical control. These results thus suggest a path towards electrically gating a variety of phenomena governed by metal/oxide interfaces, and provide a novel means to locally and reversibly imprint material properties by local activation of ionic migration.

**Methods**

**Sample preparation:** Ta(4 nm)/Pt(3 nm)/Co(0.9 nm)/GdO$_x$(3 nm) films were prepared by dc magnetron sputtering at room temperature under 3 mTorr Ar with a background pressure of $\sim 1\times 10^{-7}$ Torr, on thermally-oxidized Si(100) substrates. For the samples described in Fig. 3, the top GdO$_x$ layer was 30nm thick. All GdO$_x$ layers were deposited by reactive sputtering from a metal Gd target at an oxygen partial pressure of ~5x10$^{-5}$ Torr. Gate electrodes of GdO$_x$(30 nm)/Ta(2 nm)/Au(12 nm) were patterned using electron beam lithography and lift-off. For the samples with the 30nm thick GdO$_x$ top layer described in Fig. 3, the Ta(2 nm)/Au(12 nm) electrodes were deposited through a shadow mask.

*In situ* **high resolution TEM characterization:** Microstructural analysis and electron energy loss spectroscopy (EELS) were performed on a JEOL 2200FS TEM with double Cs correctors, operated at 200 keV. A cross-sectional TEM specimen was fabricated from a patterned Si/SiO$_2$/Ta(4 nm)/Pt(3 nm)/Co(0.9 nm)/GdO$_x$(30 nm)/Ta/Au sample using the following steps: First, a Si substrate was glued to the top surface of the sample and the Si/multilayer/Si sandwich was subsequently cut into thin slices. Next, a thin slice was polished into a wedge by a MultiPrep polishing machine (Allied High-Tech). After gluing the specimen to a half TEM Cu grid, it was further polished by Ar ion milling. Before mounting the grid onto an *in situ* electrical probing holder (HE150, Nanofactory Instruments AB), the Si was unglued from the wedge using acetone. After Si removal, a piezo-controlled Pt/Ir tip with a diameter of about 40 nm was able to contact the patterned electrode on top of the GdO$_x$ layer. Slight bending of the sample was observed after contact, but the structural integrity of the layers remained intact (Fig. 1b). Silver paste was used to make electrical contact between the bottom electrode of the layer structure and the Cu grid (see Fig. S1). The thickness of the TEM specimen was estimated to be less than 30 nm by measuring the intensity ratio of the plasmon loss and the zero-loss peaks in EELS. For the analysis of EELS core-loss peaks, background subtraction was performed using a power-law fit. The lateral resolution of STEM-EELS characterization was about 0.25 nm.

**Magneto optical Kerr effect measurements:** Polar magneto optical Kerr effect (MOKE) measurements were made using a 532 nm diode laser attenuated to 1 mW, except where noted. The laser was focused to a ~3 μm diameter probe spot and positioned by a high resolution (50 nm) scanning stage with integrated temperature control. Gate voltage was applied using a mechanically-compliant BeCu microprobe. Mechanically-generated nucleation sites created for the measurements in Fig. 2 were prepared by applying mechanical stress to the film surface using a stiff W microprobe tip. Magnetic hysteresis loops were measured at a sweep rate of 28.3 kOe/s, using an electromagnet with a risetime of ~300 μs and a maximum amplitude of 650 Oe.

The time-resolved domain expansion snapshots in Figs. 4 and 5 were obtained by, at each pixel, first saturating the magnetization and then applying a reverse field step ($H$=90 Oe in Fig. 4, and $H$=42 Oe in Fig. 5) while acquiring a time-resolved MOKE signal transient. Five reversal cycles were averaged at each pixel, from which the average trajectory of the expanding domain was reconstructed.

Time resolved MOKE transients along a line extending radially from a nucleation site were used to determine the domain wall velocity reported in Fig. 5(g). At each position, 25 reversal cycles were acquired and averaged, yielding the cumulative probability distribution of switching times. The mean reversal time $t_{1/2}$, taken as the time at which the probability of magnetization switching is 50 %, was plotted versus position, and the slope used to determine the mean velocity.

**Laser induced temperature rise:** To estimate the laser-induced temperature increase $\Delta T$, we used the temperature dependence of the coercivity of a submicrometer patterned feature. We first measured $H_c$ versus substrate temperature, using a temperature-controlled stage and a low incident laser power (<1mW) for the MOKE probe spot, and then measured $H_c$ versus incident laser power $P$, at a fixed substrate temperature. We estimate that $P = 1$ mW corresponds to a negligible $\Delta T$ whereas $P = 10$ mW corresponds to a $\Delta T$ of at least ~20 °C.

**Acknowledgement**

This work was supported by the National Science Foundation under NSF-ECCS -1128439 and through the MRSEC Program under DMR-0819792. Technical support from David Bono and Mike Tarkanian is gratefully acknowledged. Work was performed using instruments in the MIT Nanostructures Laboratory, the Scanning Electron-Beam Lithography facility at the Research Laboratory of Electronics, and the Center for Materials Science and Engineering at MIT. *In situ* TEM and EELS characterization was conducted using the facilities of the Aalto University Nanomicroscopy Center (Aalto-NMC) in Finland.

**Author Contributions**

U.B. and G.B. conceived and designed the experiments. H.L.T. proposed the extension of studies to higher temperatures. U.B. prepared the samples with help from S.E.. U.B. performed the MOKE experiments and analyzed the data. S.v.D. and L.Y. performed and analyzed the TEM and EELS measurements. U.B wrote the manuscript with assistance from G.B. and input from S.v.D. and L.Y. All authors discussed the results.

**Additional Information**

The authors declare no competing financial interests.

**Figure Captions**

**Figure 1 | Cross-sectional TEM and EELS analysis. a,** High-resolution TEM image of the SiO$_2$/Ta/Pt/Co/GdO$_x$ layer structure. STEM-EELS spectra were measured along green line, starting from the Pt/Co interface as indicated by white arrow. **b,** Contact between the Pt/Ir probe of the *in situ* TEM holder and the Ta/Au top electrode (TE). The Ta/Pt/Co bottom electrode is indicated by BE. **c,** STEM-EELS spectra of the O-K edges and the Co-L$_{2,3}$ edges from a location in the center of the Co layer before (black) and after (red) applying a negative bias of -3V. **d-f,** O K-edge and Co L$_3$-edge count rates along a similar line profile as indicated in Fig. 1**a**. The data are obtained before the application of a bias voltage (**d**) and after applying -3V (**e**) and -5V (**f**), respectively.

**Figure 2 | Device schematics and voltage control of magnetic anisotropy. a.** Topographic map of the coercivity ($H_c$) in the virgin state, in the vicinity of a gate electrode. **b.** Schematic view of gate electrode structure. **c.** and **d.** show $H_c$ and the implied magnetic anisotropy energy (MAE) landscape, respectively, after applying a gate voltage $V_g$ = -4 V for 240 s at room temperature (RT); **e.** and **f.** show the same after applying $V_g$ = -4 V for 155 s at 100 °C. **g-k,** Polar MOKE hysteresis loops measured at room temperature at the center of the gate electrode showing the device in its virgin state (**g**), after applying $V_g$ = -4 V at 100 °C for 1 s (**h**), 150 s (**i**) and 230 s (**j**), and after applying $V_g$ = +4 V at 100 °C for 270 s (**k**). Note that the Kerr signal intensity in (**i**) is reduced by a factor of 2 and in (**j**) by a factor of 16, as indicated by inset number.

**Figure 3 | Voltage-induced propagation of oxidation front. a-d**, Schematics of Pt/Co/Gd/GdO$_x$ samples with different Gd spacer thicknesses (0 to 1 nm) at the Co/GdO$_x$

interface and polar MOKE hysteresis loops corresponding to the as-deposited samples. **e-h**, Evolution of polar MOKE hysteresis loops after application of positive (**e,f**) and negative (**h**) gate voltage $V_g$ at 100 °C to the sample with 0.7 nm thick Gd spacer layer (**c,g**). All hysteresis loops were measured at room temperature (RT) and zero bias. The red curve in panel (f) shows a hysteresis loop after first setting the device in a state with minimum remanence ratio using a negative gate voltage, and then applying a 10ms voltage pulse $V_g$=+12V at 120°C, returning the device close to its initial state.

**Figure 4 | Effects of voltage and laser illumination on magnetic anisotropy. a**, Polar MOKE hysteresis loops measured inside gate electrode with the device in its virgin state (black line), after application of a gate voltage $V_g$ = -3V for 90 s (orange line) and after application of $V_g$ = -7 V for 180 s under laser illumination (green line). **b-d.** Snapshots of domain expansion around a laser-induced nucleation site, at the indicated times following application of a reverse field step of 90 Oe at t=0. All snapshots were acquired at zero bias. The dashed black line in (**b**) outlines the area exposed for 100 s to the 10 mW laser spot at a gate voltage -3 V. **e.** Schematic showing top view of an electrode in which a laser-induced nucleation site has been created at the upper-right corner, and the probe laser spot is positioned at the bottom left corner. **f,** Voltage dependence of coercivity $H_c$ as a function of $V_g$, corresponding to the schematic experiment geometry in (**e**), as $V_g$ is cycled between +/-3 V.

**Figure 5 | Laser-defined anisotropy patterns and domain wall conduits. a-f,** Time-resolved polar MOKE maps showing domain expansion in laser-defined domain wall conduit with increasing time *t* after application of a magnetic driving field *H* = 42 Oe. Dashed black lines in

(**a**) outline area exposed by laser spot with 10 mW incident power, under gate voltage $V_g$ = -3 V. The dashed circle was exposed for 100 s while dashed, L-shaped line was scanned in 1.25 µm steps with 65 s exposure at each point. All maps were acquired under zero bias. **g**. Domain wall velocity as a function of magnetic field $H$ in virgin film and along conduit exposed at 10 mW incident power under $V_g$ = -3 V for 60 s and 65 s.

**Figures:**

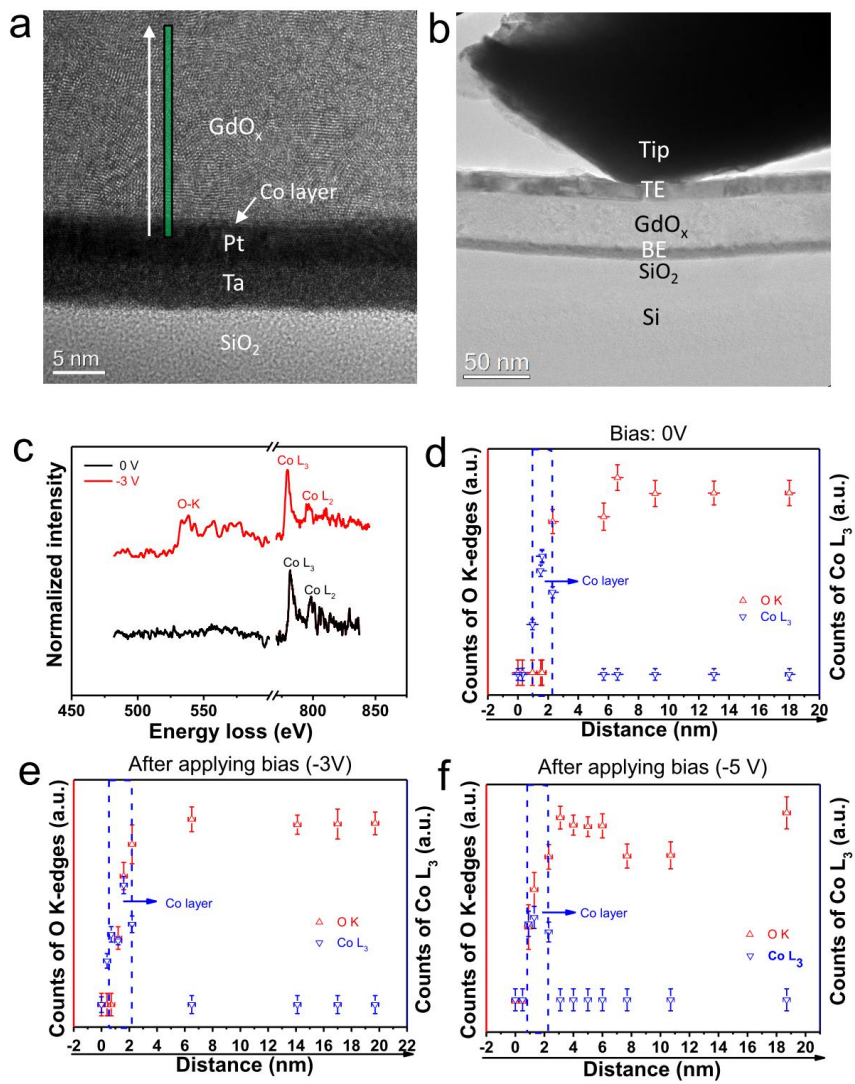

Figure 1

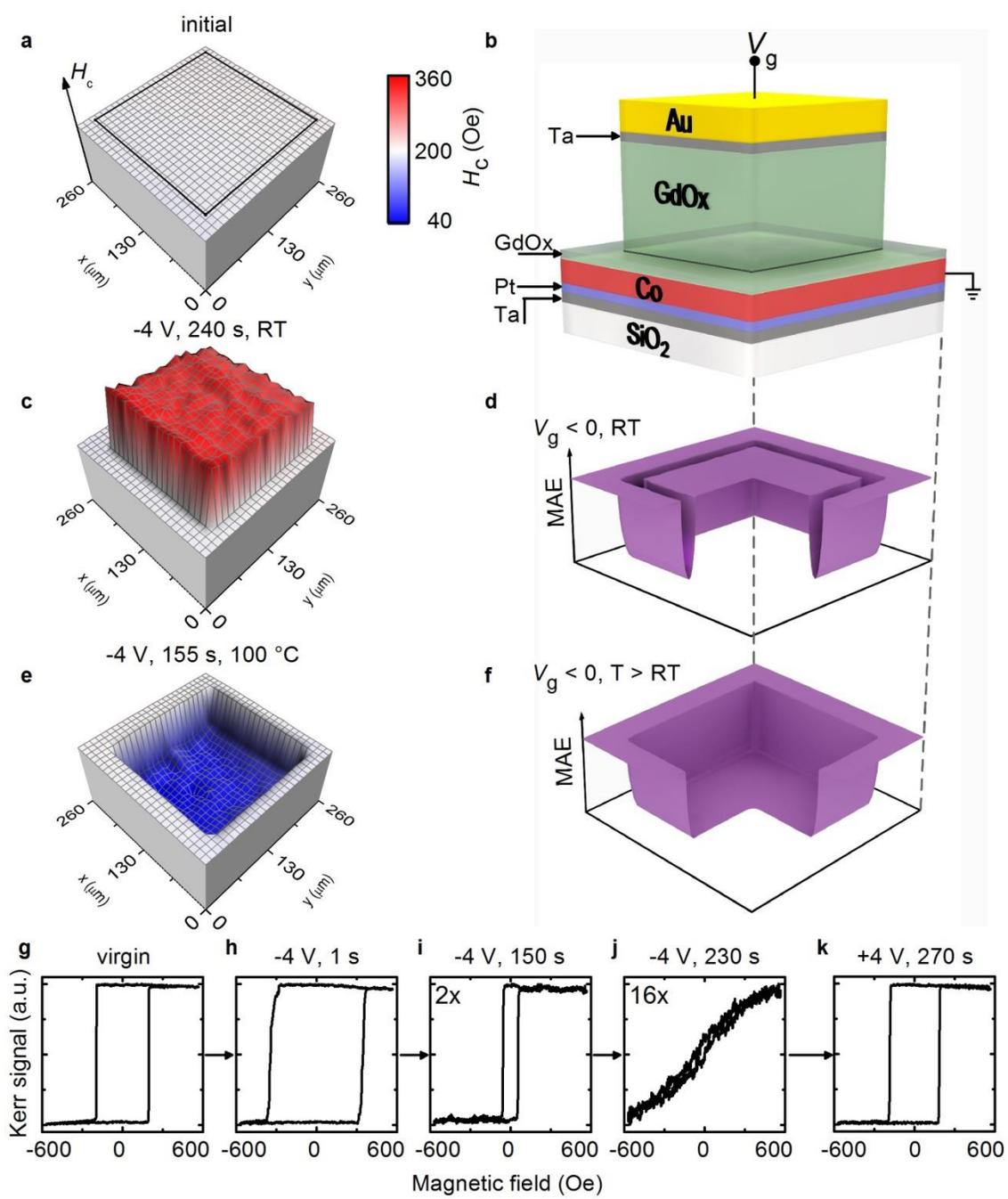

Figure 2

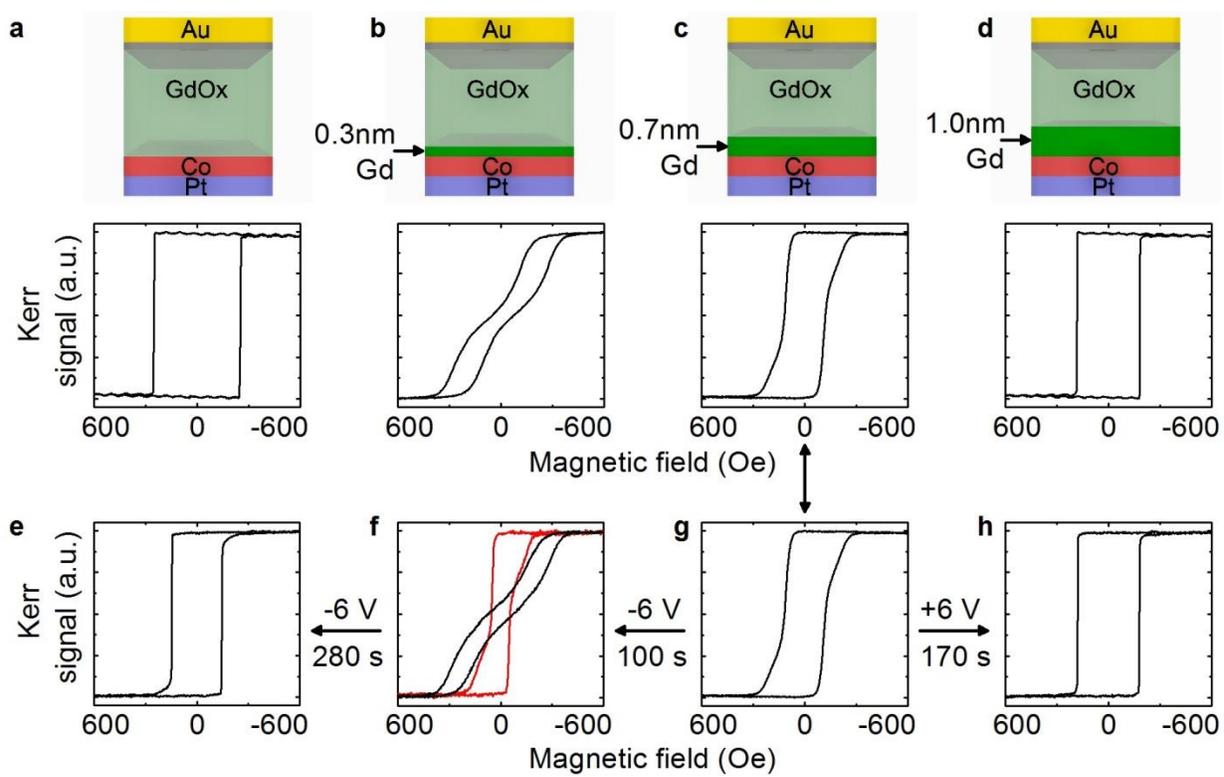

Figure 3

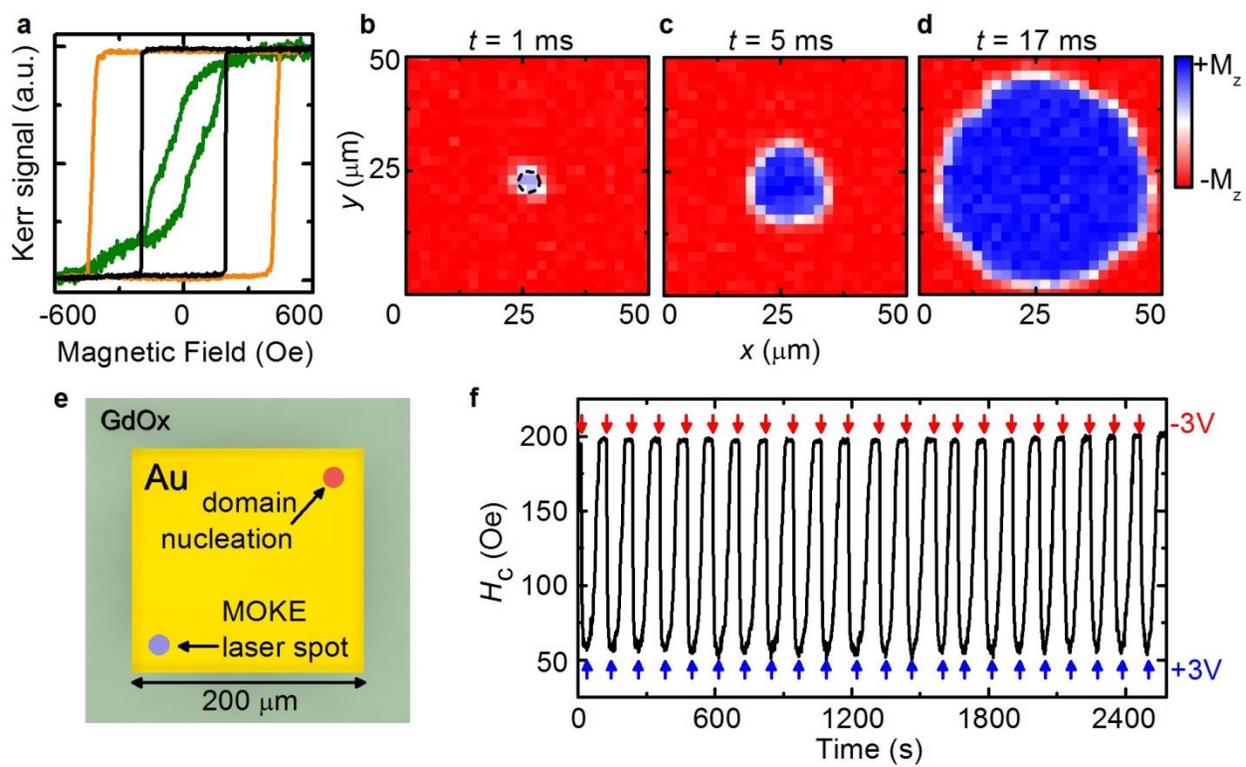

Figure 4

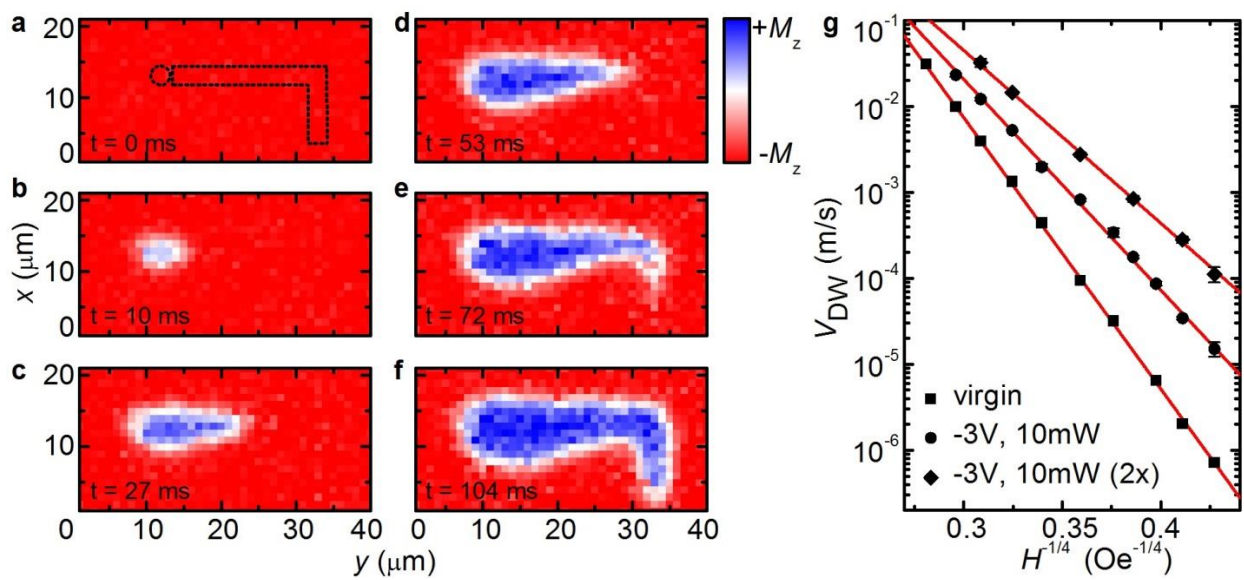

Figure 5